\documentclass[12pt]{amsart} 

\usepackage{amssymb} 
\usepackage{amsmath} 
\usepackage{fullpage} 
\usepackage{url}

\newtheorem{theo}{Theorem}

\theoremstyle{definition}
\newtheorem{defi}{Definition}
\theoremstyle{remark}
\newtheorem{rem}{Remark}

\def\Er{{\mathbb E}}

\def\Pr{{\mathbb P}}
\def\Qr{{\mathbb Q}}
\def\Rr{{\mathbb R}}

\def\Ac{{\mathcal{A}}}

\def\Fc{{\mathcal{F}}}

\def\Sc{{\mathcal{S}}}

\def\one{{\rm \bf 1}}
\def\({\left(}     
\def\){\right)}    
\def\[{\left[}     
\def\]{\right]}

\def\ep{\varepsilon}
\usepackage{color}

\begin{document}
\title{Surplus Sharing with Coherent Utility Functions} 
\author{Delia Coculescu and Freddy Delbaen}
\address{\noindent Delia Coculescu: \newline Institute for Banking and Finance, University of Zurich, Plattenstrasse 32, 8032 Zurich, Switzerland}
\email{delia.coculescu@math.uzh.ch}
\address{\noindent Freddy Delbaen: \newline Departement f\"ur Mathematik, ETH Z\"urich, R\"{a}mistrasse 101, 8092 Z\"{u}rich, Switzerland \newline Institut f\"ur Mathematik, Universit\"at Z\"urich, Winterthurerstrasse 190, 8057 Z\"urich, Switzerland}
\email{delbaen@math.ethz.ch}

\date{\today}

\begin{abstract}
We use the theory of coherent measures to look at the problem of surplus sharing in an insurance business.  The surplus share of an insured is calculated by the surplus premium in the contract. The theory of coherent risk measures and the resulting capital allocation gives a way to divide the surplus between the insured and the capital providers, i.e. the shareholders. \end{abstract}

\maketitle

\section{Notation and Motivation}

In the present work we analyse a method to distribute the surplus of an insurance business between the different agents taking part in the risk exchange.  The insured pay premia to an insurance company and in exchange of this, the company takes over the risks, i.e. the company will pay out the claims that otherwise would have to be covered by the insured.  Besides these two players, there is also the supervision  or the regulator.  The task of the regulator is to make sure that the insurance business is fair.  That means for instance that all companies play with the same rules so that competition can take place.  The regulator must also see whether the companies have enough capital to cover the risks since otherwise -- in case of bankruptcy -- a substantial portion of the risk would be transferred to society, i.e. to the tax payers or to other economic agents. 

As already pointed out by Deprez and Gerber \cite{DG}, insurance premia should be dependent on the whole portfolio of insurance contracts. That would mean that the premium to be charged for a contract can only be calculated when all other contracts (signed or in the ``pipeline") are known.  In practice this is impossible and hence premia are charged that are certainly higher than the fair allocation of the total premium to the individual contracts.  The result is that the extra part must be seen as a contribution to the capital of the company and hence is entitled to a share in the eventual surplus.

In the present document we will deal with a way to calculate the share of each economic agent, be it the capital providers or shareholders or the insured through their extra premium.  The problem on how to fix the total amount of regulatory capital  is the subject of joint ongoing work, which was initiated by Artzner and Eisele \cite{AE}.

We also restrict our analysis to a one period model.  The surplus share is particularly important for life insurance contracts and handling these would necessitate a multi-period setup.  The technical and conceptual problems are not easy so we ``postponed" it to further research.

We will use the language of probability theory as is usually done in financial and actuarial mathematics. We will fix a probability space $(\Omega, \Fc,\Pr)$ on which all random variables will be defined.  In particular the claims will be seen as random variables defined on this probability space. For simplicity we will only work with bounded random variables.  The vector space of bounded random variables is denoted by $L^\infty(\Omega,\Fc,\Pr)$ or simply $L^\infty$. The restriction to bounded random variables facilitates the modelling since we do not have to make assumptions on integrability, big tails, and so on. However it triggers some extra mathematical problems.  The solution of these problems will only be sketched and for more details we refer to \cite{Pisa} and \cite{FDbook}.  We assume that the insurance company takes decisions using a coherent utility function.
 \begin{defi} A mapping $u\colon L^\infty\rightarrow \Rr$ is called a monetary utility function if the following properties hold
 \begin{enumerate}
 \item if $0\le \xi\in L^\infty$ then $u(\xi)\ge 0$,
 \item $u$ is concave i.e. for all $\xi,\eta\in L^\infty$, $0\le \lambda\le 1$ we have $u(\lambda \xi +(1-\lambda)\eta)\ge \lambda u(\xi) + (1-\lambda) u(\eta)$,
 \item for $a\in \Rr$ and $\xi\in L^\infty$, $u(\xi +a)=u(\xi) + a$
 \item if $\xi_n\downarrow\xi$ then $u(\xi_n)\rightarrow u(\xi)$.
 \end{enumerate}
 If moreover for all $0\le \lambda\in \Rr$, $u(\lambda \xi)=\lambda u(\xi)$, we call $u$ coherent.
 \end{defi}
 
 The number $u(\xi)$ can be seen as a risk adjusted valuation of the future uncertain position $\xi$. Property (1) in the definition is therefore clear.  Risk averseness is usually translated by concavity properties and it is believed that combinations are less risky than individual positions.  This explains property (2).  Property (3) means that risk adjusted valuations are measured in money units.  Of course money today (date of the valuation) is different from money at the end of the period. Introducing a deflator or discounting -- as is the practice in actuarial business since hundreds of years -- solves this problem.  It complicates notation and as long as there is only one currency involved it does not lead to confusion if one supposes that this discounting is already incorporated in the variables. The fourth property is a continuity property.  Using monotonicity (a consequence of the previous properties, see \cite{FDbook}), we can also require that $u(\xi_n)\downarrow u(\xi)$.  In this text we will use the more stronger property where $\downarrow$'s are replaced by $\uparrow$'s, see \cite{FDbook}.  This avoids some mathematical problems that are easily overcome but they obscure the philosophy of the approach.  The homogeneity property is a strong property.  In a later paragraph where we use commonotonicity, positive homogeneity is already satisfied.
 
Sometimes the value $u(\xi)$ only depends on the distribution or law of the random variable $\xi$. In this case we say ``law invariant", ``law determined", ...  An example of such a mapping is the distorted probability.  These were introduced in insurance by Yaari, \cite{Yaa} and Denneberg \cite{Denn}.  They were later used by Wang \cite{Wang}.  An example of such a utility function is (for $\xi\ge 0$) the Choquet integral $\int_0^\infty f(\Pr[\xi>t])\,dt$ where $f:[0,1]\rightarrow [0,1]$ is convex, $f(0)=0,f(1)=1$. See \cite {FDbook}, \cite{Schm1}, \cite{Schm2}, \cite{D1} for more details and for the relations between convex games and commonotonicity.

We say that a random variable $\xi$ is acceptable if $u(\xi)\ge 0$.  Remark that $\xi-u(\xi)$ is always acceptable.  If $u$ is coherent then the acceptability set $\Ac=\{\xi\mid u(\xi)\ge 0\}$ is a convex cone.  The continuity assumption allows to apply convex duality theory and leads to the following representation theorem
\begin{theo} If $u$ is coherent there exists a convex closed set $\Sc\subset L^1$, consisting of probability measures, absolutely continuous with respect to $\Pr$, such that for all $\xi \in L^\infty$:
$$ u(\xi)=\inf_{\Qr\in\Sc} \Er_\Qr[\xi].$$
Conversely each such a set $\Sc$ defines a coherent utility function.
\end{theo}
\begin{rem}  We identify an absolutely continuous probability measure $\Qr\ll\Pr$ with its Radon-Nikodym derivative $\frac{d\Qr}{d\Pr}$.
\begin{rem} Replacing the true or physical probability $\Pr$ by other measures is a practice that is well known in insurance. Standard techniques that can be described  are e.g ``tilting" and  increasing or decreasing ages in life insurance contracts. To calculate premia, more weight is then given to unfavourable events and favourable events get less weight. The existence of such a set $\Sc$ says that the change in probability is done in a systematic and consistent way.
\end{rem}
\end{rem}
\begin{rem} We can show that the upward sequential continuity in the definition of coherent utility functions is equivalent to the existence of a minimizer $\Qr_0\in \Sc$. This equivalence follows from a deep mathematical result in functional analysis (theorem of R. James), \cite{FDbook}.  That the $\inf$ can be replaced by $\min$ simplifies some of the proofs.  We call this property the weak compactness property of $\Sc$.
\end{rem}
\begin{defi}  We say that two random variables $\xi,\eta$ are commonotonic if there exist a random variable $\zeta$ as well as two non-decreasing functions $f,g\colon \Rr\rightarrow\Rr$ such that $\xi=f(\zeta)$ and $\eta=g(\zeta)$.
\end{defi}
\begin{rem} It is a non-trivial exercise to show that we can always take $\zeta=\xi+\eta$.
\end{rem}
\begin{defi} We say that $u\colon L^\infty\rightarrow\Rr$ is commonotonic if for each couple $\xi,\eta$ of commonotonic random variables we have $u(\xi+\eta)=u(\xi)+u(\eta)$.
\end{defi}
\begin{rem}  Commonotonic concave monetary utility functions are positively homogeneous and hence coherent. In general coherent utility functions satisfy $u(\xi+\eta)\ge u(\xi)+u(\eta)$. That means that by diversifying, the risk adjusted valuation gets better. Commonotonic random variables form the opposite of diversification. Roughly speaking: what is worse for $\xi$ is worse for $\eta$. The commonotonicity of $u$ can therefore be seen as a translation of the rule:  if there is no diversification, there is also no gain in putting these claims together. 
\end{rem}

\section{Description of the Model}

We use the following setup.  There are $N$ agents to be insured, indexed $i=1,\ldots, N$.  There is one insurer denoted by the index $0$.  There is also a ``super"-reinsurer whose role will be explained later.  He will be denoted by $r$. The agents have liabilities that they want to insure.  The liability for agent $i$ is $X_i\ge 0$.  Without insurance her position will be $-X_i$. There are different premium principles which will be described in the examples below.  The utility functions of the agents are denoted by $u_i$.  The coherent utility function of the insurer is $u_0$ and to reduce complexity, supposed to be commonotonic.  The utility functions of the agents are more restrictive than the insurer's utility function $u_0$. Meaning: when a random variable is not acceptable for the insurer, then it is not acceptable for the agents. Equivalently we can say that acceptable elements for the agent $i$ are acceptable for the insurer. This is a translation of the fact that the agent $i$ feels a need for insurance and that an insurer can fill in these needs.   Because of the translation property we then have for all $\xi$ and all $i\ge 1$:
$$
u_0(\xi)\ge u_i(\xi).
$$

The insurer also brings in an initial capital $k_0$.  He will take the insurance only if he can obtain a better outcome.  Because $u_0$ is coherent the total premium, $\pi_0$, must be at least $\pi_0=-u_0\(-\sum_{i=1}^N X_i \) $.  Indeed, the insurer takes the random variable $-\sum_iX_i$ and receives $\pi_0$.  The deal is only acceptable for the insurer if $u_0(\pi_0-\sum_iX_i)\ge 0$. We suppose that the scenario set for $u_0$, $\Sc_0$  is weakly compact so that there is a $\Qr_0\in\Sc_0$ such that 
$$
\Er_{\Qr_0}\[\sum_{i=1}^N X_i\]=\sup_{\Qr\in\Sc_0}\Er_\Qr\[\sum_{i=1}^N X_i\]=-u_0\(-\sum_{i=1}^N X_i \).
$$
In case $u_0$ is commonotonic, we can (and shall) even choose $\Qr_0$ so that for all $Y$ commonotonic with $\sum_i X_i$ and such that the distribution of $Y$ has no other points of increase as $\sum_iX_i$, we have $u_0(Y)=\Er_{\Qr_0}[Y]$. This applies for cases such as $Y=\min(R,\sum_i X_i)$  or $Y=\(\sum_i X_i-R\)^+$), where $R\in \Rr$.

From \cite{Pisa}, \cite{FDbook} 
using the capital allocation principle we find that  the  individual ``fair" premia should be
\begin{equation}\label{pi}
\pi_i=\Er_{\Qr_0}[X_i].
\end{equation}
In \cite{Pisa} and \cite{FDbook} it is shown that the allocation principle can also be obtained as an application of the marginal contribution of one agent.  The latter was the approach given in \cite{DG}.  The equivalence requires some technical assumptions that go beyond the scope of this paper. What it means is that under these extra assumptions we have
$$
\pi_i=\lim_{\ep\downarrow 0}(-1)\frac{u_0(-\sum_jX_j-\ep X_i)-u_0(-\sum_jX_j)}{\ep}
$$
The equality $\pi_i=\Er_{\Qr_0}[X_i]$ immediately implies $$-\pi_i =\Er_{\Qr_0}[-X_i]\ge u_0(-X_i)\ge u_i(-X_i).$$ 
For the agent $i$ this is a good deal, provided that the insurance pays entirely the claim $X_i$ at date 1,  since $-\pi_i\ge u_i(-X_i)$. What she pays for insurance is better than paying $X_i$.

In the following, we propose four different models, that is, four different ways in which the premia can be fixed by the insurer in situations where there is either a government guarantee, or a reinsurance possibility. There is a distinction between the fair premia and what is actually charged, which may be higher. We argue that the difference should be regarded as a participation to the capital by the insured agents, and be remunerated by a share of the surplus. We provide conditions that the deals are acceptable by all agents and the insurer. More precisely, we consider  that a deal is acceptable for the insurer whenever the utility $u_0$ of the profit received by the insurer exceeds $k_0$. We say that the deal is acceptable for agent $i$ whenever the insurance deal generates an increase in agent's utility $u_i$, as compared with the situation without insurance.

\section{Model 1}

In this example we suppose that the insurer has limited liability.  We take for the total premium $\pi_0=-u_0\(-\sum_{i=1}^N X_i \)$. We distinguish several cases:
\begin{enumerate}
\item $\sum_i X_i > \pi_0 +k_0$.  In this case the total claim size exceeds the available capital.  The excess is supposed to be covered by for instance the government and this at no cost.  The initial capital should then be sufficiently high to make the deal acceptable for the government.  The determination of this level is beyond the contents of this paper.  We denote by $A$ the set $A=\{\sum_i X_i > \pi_0 +k_0\}$.
\item $\pi_0 \le \sum_i X_i \le \pi_0 +k_0$.  In this case there is no surplus and the insurer will lose part of his investment.  We denote by $B$ the set $B=\{\pi_0 \le \sum_i X_i \le \pi_0 +k_0\}$.
\item $\pi_0 > \sum_i X_i $.  In this case there is a surplus.  The insurer will keep the surplus entirely.  This can be defended since the agents already ``gained" from the allocation principle which is their share when entering the insurance. Also they do not take any risk. We denote by $C$ the set $C=\{\pi_0 > \sum_i X_i \}$.
\end{enumerate}
\begin{theo} The deal is acceptable for the insurer.
\end{theo}   
{\bf Proof}\quad The insurer will accept the deal if
$$
u_0(\one_{B\cup C} (\pi_0+k_0-\sum_i X_i ))\ge k_0.
$$
This is easily proved.  By definition of $\pi_0$, we have $u_0(\pi_0-\sum_i X_i)=0$, hence $u_0(\pi_0-\sum_i X_i+k_0)=k_0$, therefore by monotonicity $u_0\((\pi_0-\sum_i X_i+k_0)^+\)\ge k_0$.\qed
\begin{rem}  We remark that there must be a regulator who requires a minimum  capital $k_0$.  Otherwise the company would choose $k_0=0$, take the profit on the event $C$ and leave the trouble to ``society" on the event $A$.   As said in the introduction,  the rules used by the regulators  and the implementation in models is the topic of research with Artzner and Eisele. 
\end{rem}
\section{Model 2}

This is almost the same as example 1, but this time we require a premium for covering the excess. We also assume that the reinsurer has no default. The reinsurance premium will be calculated by the same coherent utility function, i.e. the same set $\Sc_0$.  It is here that we use the commonotonicity.  The retention will be denoted by $R$ and this results in the splitting:
$$
\sum_i X_i=\(\sum_i X_i\)\wedge R + \(\sum_i X_i-R\)^+.
$$
The two terms are commonotonic and hence the premium satisfies
$$
\pi_0=\pi^R + \rho^R;\quad \pi^R=\Er_{\Qr_0}\[\(\sum_i X_i\)\wedge R\];\quad \rho^R=\Er_{\Qr_0}\[ \(\sum_i X_i-R\)^+\].
$$
  The retention level needs to be chosen so that the claims are fully covered, that is, the available funds (given by $ k_0+\pi^R$)  are not below the retention level. We define the optimal retention $R$ as the maximal retention level that results in a full coverage of the claims:
$$
R:= \max\{x\;|\; x\leq  k_0+\pi^R\}.
$$
The existence and uniqueness of $R$ follows from an easy analysis of the function $\Phi$ defined as
$$
\Rr_+\rightarrow \Rr_+;\;  x\rightarrow x-\Er_{\Qr_0}\[\(\sum_i X_i\)\wedge x\].
$$
As $x-\Er_{\Qr_0}\[\(\sum_i X_i\)\wedge x\]=\Er_{\Qr_0}\[\(x-\sum_i X_i\)^+\]=\int_0^x \Qr_0\(\sum_i X_i\leq a\)da$, the function $\Phi(x)$ is continuous, convex, strictly increasing after it leaves zero, is $0$ at $0$ and tends to $\infty$ for $x\rightarrow +\infty$.

Hence, the optimal retention $R$ satisfies:
$$
R=  k_0+\pi^R,
$$that is, the available capital is $R$, which is also the maximum the insurer has to pay out. The surplus is therefore
$$
R-\(\sum_i X_i\)\wedge R.
$$
The agents again do not take any risk and hence they should not participate in the surplus.  The insurer finds it a good deal if
$$
u_0\(R-\(\sum_i X_i\)\wedge R\)\ge k_0.
$$
But the definition of $R$ shows that $u_0\(R-\(\sum_i X_i\)\wedge R\) = k_0$.  That means there is no incentive to do business and the insurer must get all the profit to have an equivalent outcome.
\section{Model 3}
This is an extension of the previous models.  The agents pay a premium equal to $p_i\ge \pi_i$, where $\pi_i$ is the fair premium, introduced in (\ref{pi}). This has the advantage that the insurer can announce the premium without having to calculate the total premium necessary to cover the total losses.  Of course this procedure should lead to a premium greater than the fair premium as calculated in the previous models. An example of such a premium calculation could be the amount $\sup_{\Qr\in \Sc_0}\Er_\Qr[X_i]=-u_0(-X_i)$. The extra premium $p_i-\pi_i$ can be seen as a contribution of agent $i$ to the capital, hence agents should be entitled to a share in the surplus. All investors must be treated in the same way and hence the share of agent $i$ is proportional to her contribution, namely $\frac{p_i-\pi_i}{\sum_j (p_j -\pi_j)+ k_0}$.  The investor will get a proportion $\frac{k_0}{\sum_j (p_j -\pi_j)+ k_0}$.  These fractions are paid out regardless of having caused a claim or not.

The retention is now defined by the relation
$$
\sum_i(p_i-\pi_i) + k_0 +\pi^R=R.
$$
The existence and uniqueness of $R$ are proved in the same way. This time we must see whether this is a good deal for the insurer as well as for the agents.
\begin{theo} The deal is acceptable for the insurer.\end{theo}
{\bf Proof}\quad   For the insurer we must check the inequality
$$
u_0\(\frac{k_0}{k_0+\sum_i\(p_i-\pi_i\)}\(R-\sum_iX_i\)^+\)\ge k_0.
$$
By positive homogeneity of $u_0$ this is the same as:
$$
u_0\( R-\(\sum_i X_i\)^+\) \ge k_0+\sum_i\(p_i-\pi_i\).
$$
As in the previous example this follows from the definition of $R$ which implies:
$$
u_0\( R-\(\sum_i X_i\)\wedge R\) = R-\pi^R= k_0+\sum_i\(p_i-\pi_i\).
$$
\qed
\begin{theo} The deal is acceptable for the insured as soon as $p_i\le \sup_{\Qr\in\Sc_0}\Er_\Qr[X_i]$.
\end{theo}
{\bf Proof}\quad For agent $i$ we must check:
$$
u_i\(-p_i+\frac{p_i-\pi_i}{k_0+\sum_j\(p_j-\pi_j\)}\(R-\sum_iX_i\)^+\)\ge u_i\( -X_i\).
$$
This is equivalent to
$$
u_i\(\frac{p_i-\pi_i}{k_0+\sum_j\(p_j-\pi_j\)}\(R-\sum_iX_i\)^+\)\ge u_i\( p_i-X_i\).
$$
The left hand side is positive whereas the right hand side is negative provided the premium $p_i$ is not too big.  For instance if $p_i\le  \sup_{\Qr\in\Sc_0}\Er_\Qr[X_i]$ we have $u_0\(p_i-X_i\)\le 0$ and hence also $u_i\(p_i-X_i\)\le 0$. \qed
\begin{rem} In any case the agent $i$ will not pay a premium $p_i$ that is bigger than $-u_i(-X_i)$.  Paying a higher premium and counting on surplus participation is not realistic since the  surplus share also depends  on the claims incurred through the other agents.
\end{rem}

\section{Model 4}

We continue the building of stepwise more complicated models.  We suppose that there are two insurers.  The first one has a utility function $u_0$, described by the scenario set $\Sc_0$.  This insurer acts as the direct insurer.  The second insurer acts as a reinsurer with utility function $u_r$ described by the scenario-set $\Sc_r$. For a claim $0\le \xi$, the reinsurer would charge a premium $\sup_{\Qr\in\Sc_r}\Er_\Qr[\xi]$.

Both utility functions $u_0$ and $u_r$ are supposed to be commonotone.  Their scenario-sets are therefore determined as cores of convex games, say $v_0\ge v_r$. 
For simplicity we suppose that both scenario-sets $\Sc_0, \Sc_r$ are weakly compact.  We can therefore suppose that there are elements $\Qr_0\in\Sc_0,\Qr_r\in\Sc_r$ such that for all $a$:
\begin{align*}
\sup_{\Qr\in\Sc_r}\Er_\Qr\[\sum_jX_j\]&=\Er_{\Qr_r}\[\sum_jX_j\]\\
\sup_{\Qr\in\Sc_r}\Er_\Qr\[\(\sum_jX_j\)\wedge a\]&=\Er_{\Qr_r}\[\(\sum_jX_j\)\wedge a\] \\
\sup_{\Qr\in\Sc_r}\Er_\Qr\[\(\sum_jX_j-a\)^+\]&=\Er_{\Qr_r}\[\(\sum_jX_j-a\)^+\] \\
\sup_{\Qr\in\Sc_0}\Er_\Qr\[\sum_jX_j\]&=\Er_{\Qr_0}\[\sum_jX_j\]\\
\sup_{\Qr\in\Sc_0}\Er_\Qr\[\(\sum_jX_j\)\wedge a\]&=\Er_{\Qr_0}\[\(\sum_jX_j\)\wedge a\] \\
\sup_{\Qr\in\Sc_0}\Er_\Qr\[\(\sum_jX_j-a\)^+\]&=\Er_{\Qr_0}\[\(\sum_jX_j-a\)^+\].
\end{align*}

 The reinsurer is supposed to be default free (it could be a government institution or a solidarity fund of the insurance industry).  We suppose that the agents have access to this reinsurer.  We assume that $u_i\le u_r\le u_0$ for all $i$, or what is the same $\Sc_0\subset \Sc_r\subset \Sc_i$. This reflects the fact that on one hand the agents are more risk adverse than the insurers, and on the other hand the reinsurer being  default free, the premia for the reinsurer are higher than for the direct insurer.  
 
 The actions of the direct insurer are subject to the rule that after reinsurance, the claims must be covered completely. The direct insurer will therefore take a reinsurance with a retention $R$, to be determined later.  For the retention level $R$, the cost of the reinsurance is given by (using the commonitonicity of $u_r$):
$$
\rho^R:= -u_r\(-\(\sum_j X_j-R\)^+\)=\Er_{\Qr_r}\[\(\sum_j X_j-R\)^+\].
$$

To get full coverage of their risks without input of any capital (that is, $k_0=R=0$), the agents should pay a total premium 
$$
\rho^0=-u_r\(-\sum_j X_j\)=\Er_{\Qr_r}\[\sum_j X_j\],
$$as the direct insurer can  transfer the totality of the claims to the reinsurer at the cost $\rho^0$. Using the capital allocation principle, we deduce that the premium for such a coverage to be paid by agent $j$ is 
$$
\pi^r_j:=\Er_{\Qr_r}\[X_j\].$$
 For the moment  we consider the amounts $\pi^r_j$ as the premium-input of the agent $j$; if they pay higher premia, $p_j$, the differences, $p_j-\pi^r_j$, are regarded as capital-input.
  In practice however, the direct insurer brings in a capital $k_0>0$ (again, we consider this being fixed by a regulator), and hence the retention $R$  will not be zero. This means that whenever $R>0$, the premia $\pi^r_j$ that we are considering are not exactly the fair premia (see also Remark \ref{rem} below). 
  
  As the insurer participates with a capital $k_0$, we must determine the retention level $R$; as before we consider $R$ to be the maximal retention level  that results in a full coverage.  The full coverage condition is now given by: $R\le k_0+\sum_j p_j -\rho^R$, so that the optimal  retention is:
$$
R=\max\{x\;|\; x\leq  k_0+\sum_j p_j -\rho^R\}.
$$
The full coverage condition can be written as: $ R-\Er_{\Qr_r}\[\(\sum_j X_j\)\wedge R\] \leq k_0+\sum_j (p_j-\pi^r_j) $.
Therefore, the existence and unicity of optimal retention comes  using identical arguments as before for $\Phi$, but this time for the function  $\Psi$ defined as:
$$
\Rr_+\rightarrow \Rr_+; \; x\rightarrow x-\Er_{\Qr_r}\[\(\sum_j X_j\)\wedge x\].
$$
From now on, we always assume $R$ to be the optimal retention, that is: $R=  k_0+\sum_j p_j -\rho^R$, or, alternatively:
$$
\Psi(R) =  k_0+\sum_j (p_j-\pi^r_j).
$$ In Remark \ref{rem9} we analyse the dependence of the optimal reserve on the level of the capital.

With the capital input of the insurer being $k_0$ and the retention level $R$, the outcome (i.e., surplus) at the end of the contract is
$$
S:=k_0+\sum_j p_j-\rho^R - \(\sum_j X_j\)\wedge R\geq 0 .
$$
We denote by $\lambda_0=\frac{k_0}{k_0+\sum_j (p_j-\pi^r_j)}$ the proportion of the surplus that the direct insurer is keeping and by $\lambda_i=\frac{p_i-\pi^r_i}{k_0+\sum_j (p_j-\pi^r_j)}$ the proportion of the surplus that  agent $i$ is keeping.

 We now prove that this procedure of surplus sharing is beneficial for the insurer and for the insured. Let us fist check the utility for the direct insurer.

\begin{theo} 
The deal is acceptable for the direct insurer.
\end{theo}

{\bf Proof} \quad
For the direct insurer, the deal is acceptable if and only if the utility of his share of the surplus is not less that $k_0$, that is $u_0(\lambda_0 S)\geq k_0$. The utility of his share is given by
$$
u_0(\lambda_0 S)=\lambda_0 \(k_0+\sum_j p_j  -\rho^R -\Er_{\Qr_0}\[\(\sum_j X_j\)\wedge R\]  \).
$$
This quantity is bigger than $k_0$ if and only if
$$
-\Er_{\Qr_0}\[\(\sum_j X_j\)\wedge R\]  -\Er_{\Qr_r}\[\(\sum_j X_j-R\)^+\]\ge \sum_j (-\pi^r_j) =-\Er_{\Qr_r}\[\sum_jX_j\]=u_r\(-\sum_j X_j\).
$$
The utility functions $u_0$ and $u_r$ are commonotone and therefore we have that 
\begin{align*}
&u_0\(-\(\sum_j X_j\)\wedge R\)=- \Er_{\Qr_0}\[\(\sum_j X_j\)\wedge R\]\\
&u_r\(-\(\sum_j X_j-R\)^+\)=-\Er_{\Qr_r}\[\(\sum_j X_j-R\)^+\].
\end{align*}
Therefore the inequality is obvious, given $u_0\ge u_r$. 
\qed

\begin{theo} The deal is acceptable for the insured as soon as $p_i\le \sup_{\Qr\in\Sc_r}\Er_\Qr[X_i]$.
\end{theo}
{\bf Proof} 
\quad
To check the advantage for the insured $i$, we must show  that $u_i(\lambda_i S -p_i)\geq u_i(-X_i)$, that is:
$$
\lambda_i  u_i(S)\geq p_i+u_i(-X_i)
$$
The surplus being nonnegative, the left hand side is nonnegative. The right side is negative provided the premium $p_i$ is not too big.  For instance if $p_i\le  \sup_{\Qr\in\Sc_r}\Er_\Qr[X_i]$ we have $u_r\(p_i-X_i\)\le 0$ and hence also $u_i\(p_i-X_i\)\le 0$. \qed

\bigskip

\begin{rem}\label{rem} For each agent, the cost $p_i$ was split in two parts: premium-input $\pi^r_i$ and capital-input $p_i-\pi^r_i$, with  all premia-input summing up to $\Er_{\Qr_r}[\sum_jX_j]=-u_r\(-\sum_j X_j\)$.  This is higher than the total required premium, which should be the cost of the reinsurance contract plus the cost of the direct insurance of the claims up to the retention level, given as: 
\begin{align*}
-u_r&\(-\(\sum_jX_j-R\)^+\) -u_0\(-\(\sum_jX_j\)\wedge R\)\\
&=\Er_{\Qr_r}\[\(\sum_jX_j-R\)^+\] +\Er_{\Qr_0}\[\(\sum_jX_j\)\wedge R\].
\end{align*}Indeed, from our assumptions on the utility functions, we have
$$
\Er_{\Qr_r}\[\(\sum_jX_j-R\)^+\] +\Er_{\Qr_0}\[\(\sum_jX_j\)\wedge R\]\le \Er_{\Qr_r}\[\sum_jX_j\].
$$

Probably a more fair way to split the cost $p_i$ is to consider 
$$
\pi_i=\Er_{\Qr_r}\[X_i\one_{\{ \sum_j  X_j> R  \}}\]+\Er_{\Qr_0}\[X_i\one_{\{ \sum_j  X_j\leq R  \}}\]
$$ as being the premium-input of agent $i$, and the remaining $p_i-\pi_i$ as being the capital-input of agent $i$ (provided this is positive).
This different way of splitting the cost does not affect (of course) the way to compute the optimal retention level, nor the surplus $S$ available after payment of all claims. It is only meant to provide an alternative rule for sharing the surplus: now $\frac{k_0}{k_0+\sum_j (p_j-\pi_j)}$ is the proportion of the surplus that the direct insurer is keeping and $\frac{p_i-\pi_i}{k_0+\sum_j (p_j-\pi_j)}$ the proportion of the surplus that  agent $i$ is keeping. 
Nevertheless, for such allocations of the surplus, we did not find a simple condition for the premia $p_i$  that ensures that the deals are acceptable for the agents, while respecting $p_i\geq \pi_i$. By a simple rule, we mean a condition on all $p_i$ that does not involve agent's utility directly.

But  the reader can easily check that for the direct insurer any allocation among the insured agents of the quantity $\Er_{\Qr_r}\[\(\sum_jX_j-R\)^+\] +\Er_{\Qr_0}\[\(\sum_jX_j\)\wedge R\]$  leaves the insurer with an outcome that in $u_0$ utility is equivalent to the initial capital $k_0$.  This is even independent of the retention limit.
\end{rem}

\begin{rem}\label{rem9} We will now show that with increasing capital $k_0$, the utility for the direct insurer goes up. Let us denote by $R(k_0)$ the solution of 

$$
R-\Er_{\Qr_r}\[\(\sum_j X_j\)\wedge R\] =  k_0+\sum_j (p_j-\pi^r_j).
$$
We recall that the function $\Psi(R)= R-\Er_{\Qr_r}\[\(\sum_j X_j\)\wedge R\] $ satisfies $\Psi(0)=0$ and is continuous with $\Psi'(R)= \Qr\(\sum_j X_j\leq R\)$. Hence, it is strictly increasing after it leaves zero and tends to $\infty$ when $R$ tends to $\infty$. Also, it is convex. These properties ensure that $\Psi^{-1}$ is well defined on $(0,\infty)$ and is continuous, strictly increasing and concave.

Also, we recall that we consider that all $p_j\geq  \pi^r_j$ so that $\sum_j (p_j-\pi^r_j)\geq 0$.  We can consider $k_0> 0$, to ensure positivity of the quantity $k_0+\sum_j (p_j-\pi^r_j)$.
The implicit function theorem then shows that $\frac{dR}{dk_0}\ge 0$. 

The utility of the insurer's share of the surplus $u_0(\lambda_0S)$ can be written as:
\begin{align*}
\lambda_0& \(k_0+\sum_j p_j -\Er_{\Qr_0}\[\(\sum_j X_j\)\wedge R\]  -\Er_{\Qr_r}\[\(\sum_j X_j-R\)^+\]\)\\
&=\frac{k_0}{k_0+\sum_j (p_j-\pi^r_j)}\(k_0+\sum_j  (p_j-\pi^r_j) -\Er_{\Qr_r}\[\(\sum_j X_j\)\wedge R\]  -\Er_{\Qr_0}\[\(\sum_j X_j\)\wedge R\]\)\\
&=\frac{k_0}{k_0+\sum_j (p_j-\pi^r_j)}\( R-\Er_{\Qr_0}\[\(\sum_j X_j\)\wedge R\]\)\\
&=\frac{k_0}{k_0+\sum_j (p_j-\pi^r_j)}\Er_{\Qr_0}\[\(R-\sum_j X_j\)^+\].
\end{align*}
The first factor is clearly increasing in $k_0$ and the second factor is increasing in $R$, hence also in $k_0$.

 More important is the difference between this utility and the initial capital $k_0$:
\begin{align*}
 u_0(\lambda_0S)- k_0&=\frac{k_0}{k_0+\sum_j (p_j-\pi^r_j)}\( R-\Er_{\Qr_0}\[\(\sum_j X_j\)\wedge R\]\) - k_0\\
&=k_0 \(\frac{R -\Er_{\Qr_0}\[\( \sum_j X_j\)\wedge R\]}{k_0 +\sum_j p_j -\Er_{\Qr_r}\[\sum_j X_j\]}-1\)\\
&=k_0\( \frac{ \Er_{\Qr_r}\[\(\sum_j X_j\)\wedge R \] -\Er_{\Qr_0}\[\(\sum_j X_j\)\wedge R\]}{k_0 +\sum_j p_j -\Er_{\Qr_r}\[\sum_j X_j\]}  \).
\end{align*}
This shows that when $k_0\rightarrow \infty$ the extra return tends to $\Er_{\Qr_r}\[\sum_j X_j \] -\Er_{\Qr_0}\[\sum_j X_j\]$, but also shows that the return on the initial capital tends to zero.

\end{rem}
\section{Discussion of the models}

There are different shortcomings of the models used.  The premia paid by the agents are augmented by the administration or handling costs.  Here we might argue that in case of a claim, the agents incur these costs themselves.  In the handling costs there is also included the commission paid out to the intermediaries, the brokers. These commissions should not be too high since otherwise the agents could keep their liabilities or take insurance only for the larger part of the claims, the so called tail of the distribution.

The utility functions of the agents cannot be supposed to be positively homogeneous.  they should be concave.  However if commonotonicity is used we must suppose that the utility functions are positively homogeneous as this is a consequence of commonotonicity, \cite{FDbook}. We only needed commonotonicity for the utility function of the insurer and the reinsurer. The inequalities $u_i\le u_r\le u_0$ can be verified in case $u_0$ and $u_r$ are positively homogeneous and $u_i$ is only concave. However the inequality implies some geometric restrictions on the acceptability set of agent $i$. For instance the function $u_i$ cannot be differentiable at $0\in L^\infty$.

We supposed the presence of a reinsurer who is default free.  In some cases a government can provide a guarantee but in general we must include the possibility of default of the reinsurer.  Some countries require guarantees from the reinsurer either under the form of deposits or under the form of letters of credit issued by ``bona fide" financial institutions. 

In case of liability insurance there might be a rule that in case of default of the insurer and of the reinsurer, the agent is not liable for the remaining losses.  Especially in high end insurance contracts such rules and exceptions make the modelling extremely difficult.

In our models there is a reward for those who take the risk.  This is in contradiction with some life insurance practice where only the amount of the total premium is important.

\end{document}